\documentstyle[prc,epsfig,aps]{revtex}

\newcommand{\gap}{\mathrel{ \rlap{\raise.5ex\hbox{$>$}}
                      {\lower.5ex\hbox{$\sim$}}  } }
\newcommand{\lap}{\mathrel{ \rlap{\raise.5ex\hbox{$<$}}
                                      {\lower.5ex\hbox{$\sim$}}  } }
				      
\def\bild#1\over#2{\mathrel{\mathop{\kern5pt #1}\limits_{#2}}}
\newcommand{\ds}{\displaystyle}

\newcommand{\be}{\begin{equation}}
\newcommand{\ee}{\end{equation}}
\newcommand{\ba}{ \begin{eqnarray} }
\newcommand{\ea}{ \end{eqnarray} }
\newcommand{\bc}{\begin{center}}
\newcommand{\ec}{\end{center}}
\newcommand{\ca}{$^{14}$C\ }
\newcommand{\bea}{$^{14}$Be\ }
\newcommand{\bec}{$^{12}$Be\ }
\newcommand{\beb}{$^{10}$Be\ }
\newcommand{\li}{$^{11}$Li\ }
\newcommand{\lic}{$^{9}$Li\ }
 \newcommand{\cc}{$^{12}$C\ }
\newcommand{\cb}{$^{10}$C\ }
\newcommand{\plb}[1]{Phys. Lett. B {\bf #1}}
\newcommand{\zpb}[1]{Zeit. Phys. Lett. {\bf #1}}
\newcommand{\prla}[1]{Phys. Rev. Lett. {\bf #1}}
\newcommand{\pr}[1]{Phys. Rev. C {\bf #1}}
\newcommand{\praa}[1]{Phys. Rev. A {\bf #1}}
\newcommand{\apb}[1]{Ann. of Phys. (N.Y.) {\bf #1} }
\newcommand{\npb}[1]{Nucl. Phys. {\bf #1}}

\begin{document}
\title{\large
\bf{Two-body correlations in N=8 and 10 nuclei and effective neutron-neutron 
interactions in Tamm-Dancoff
and two-particle RPA models.}}
\author{\normalsize J.C. Pacheco
  \\
{\normalsize Departamento de F\'{\i}sica Aplicada e 
  Instituto de F\'{\i}sica Corpuscular} \\
{\normalsize Universidad de Valencia,Burjasot, Valencia, Spain}  \\
\normalsize  N.Vinh Mau  \\ 
{
\normalsize 
  Institut de Physique Nucl\'eaire,} \\
{\normalsize  F-91406 , Orsay Cedex, France } \\
}


\maketitle 

\begin{abstract}
We apply a particle-particle RPA model to study the properties of the
two-neutron valence wave function in nuclei $^{14}$C, $^{12}$Be, $^{11}$Li and
$^{14}$Be. The RPA model takes account of
two-body correlations in the cores so that it gives a better description
of energies and amplitudes than
models which assume a neutron closed shell (or subshell) core.
With a Gogny neutron-neutron effective interaction or with the equivalent
density dependent delta force we are able to reproduce the two-neutron
separation energies in these nuclei and in the corresponding cores,
except for $^{9}$Li. These calculations suggest the same 2s-1p$_{1/2}$
shells inversion in $^{12}$Be-$^{13}$Be than in $^{11}$Be.
\end{abstract}

 \section{Introduction}
In an earlier work \cite{pvm}, we calculated the properties of the two-neutron 
valence pair in the nuclei $^{11}$Li, $^{12}$Be, $^{14}$C using a Tamm-Dancoff
model and assuming the core to have a closed 1p$_{3/2}$ neutron shell.
Important ingredients in the calculations are the assumed pairing
interaction and the role of the p$_{1/2}$-s$_{1/2}$ shells
inversion that is visible in the spectrum of $^{11}$Be.

While the model  gave a coherent and reasonable description of 
these nuclei
it fails to describe $^{6}$He (taking now an alpha-particle core)
giving  a two-neutron
separation energy of several MeV instead of the experimental value of 0.97
MeV \cite{wa}. This has already been found by Esbensen et al. \cite{hb}.
When one compares $^{6}$He
described as an alpha-particle  + two neutrons to \ca for example, one sees
immediately an important difference between the two systems: for $^{6}$He the
core is well described in an Hartree-Fock model as a
pure (1s)$^2_{\nu}$(1s)$^2_{\pi}$ configuration \cite{ca}
while we have long known \cite{ck} that
in \ca the core of \cc is a mixture of states with neutrons in the
1p$_{3/2}$ or 1p$_{1/2}$ states, what is not taken into account in
Tamm-Dancoff models.

In this work, we will include the core correlations using the two-particle
RPA theory and also examine more broadly the dependence on the assumed
effective two-neutron  force.  
Zero range residual forces are very simple to use but are quite arbitrary
and binding energies are not sufficient to fix uniquely the force. 
In a recent work Garrido et al. \cite{gs} have  searched for a zero range 
density dependent force equivalent to the finite range 
Gogny interaction \cite{gb,jf} (see also \cite{be}). 
The authors fit the
force in order to reproduce the gap  calculated in nuclear matter with 
the Gogny
force which has good pairing properties in finite nuclei.  A fit to the whole
 domain of k$_F$ values determines unambigously the parameters of the
force and tells what is the cut-off on neutron energy to be used. This last
information is very important since a zero range interaction has no natural
cut-off. The density independent part of the force
reproduces the low energy properties of a free neutron-neutron system so
that their force is equivalent to a realistic effective interaction in
two-neutron and infinite systems. We will use this force in our problem and
will discuss the results compared to the zero range pairing forces used in
Tamm-Dancoff models.

In section II we briefly report on the results obtained in 
a pairing or Tamm-Dancoff model with
three effective neutron-neutron interactions. In section III we recall
 the properties and equations of the particle-particle RPA model. The
results of this model are presented and discussed in section IV for 
$^{14}$C-$^{10}$C, $^{12}$Be-$^{8}$Be and $^{11}$Li-$^{7}$Li and in section V 
for $^{14}$Be-$^{10}$Be with a discussion on the
$^{13}$Be and $^{11}$Be spectra. At the end, section VI is devoted to our
conclusions.

\section{Hamiltonian  and effective interactions}
We first make a pairing, or Tamm-Dancoff, approximation to describe core + two
neutron systems. Assuming an inert and closed
sub-shell core for neutrons we 
diagonalise the two-neutron hamiltonian in a two-neutron subspace built on
non occupied neutron states in the core:
\ba
 H_{2n} &=&\frac{{\bf p}_1^2}{2m}+\frac{{\bf
    p}_2^2}{2m}+V_{nc}(1)+V_{nc}(2)+V_{nn}(1,2)+{\ds \frac{({\bf p}_1+{\bf
    p}_2)^2}{2 A_c m}}\\
      &=&h_{nc}(1)+h_{nc}(2)+V_{nn}(1,2)+{\ds \frac{{\bf p}_1.{\bf p}_2}{A_c
	  m}}
\ea 
where the one-neutron hamiltonian is:
\ba
h_{nc}(i)&=&{\ds \frac{{\bf p}_i^2}{2 \mu}}+V_{nc}(i)\\
V_{nc}(r)&=&-V_{NZ}\left (f(r)-0.44r_0^2({\bf {l.s}})\frac{1}{r}
\frac{df(r)}{dr}\right)+
16a^2 \alpha_n\;\left(\frac{df(r)}{dr}\right)^2 \\ 
f(r)&=&\left(1+exp(\frac{r-R_0}{a})\right)^{-1}\\
V_{NZ}&=&U_0-U_{\tau} \frac{N-Z}{A_c}
\ea
$\mu$ is the reduced mass equal to ${\ds\frac{A_c\;m}{A_c+1}}$; $A_c$, N, Z are
respectively the
mass, neutron and proton numbers in the core; $R_0=r_0A_c^{1/3}$ with
$r_0$=1.27 fm, $a$=0.75 fm; the strengths    $U_{0}$ and $U_{\tau}$ are
taken the same as
in our previous papers \cite{pvm,sl}. The last term of potential, eq.(4), 
simulates
particle-phonon couplings contribution to the one-body potential \cite{nvm}.
The strengths $\alpha_n$ are  fitted in each nucleus to reproduce the
experimental  1p$_{1/2}$, 2s, 1d$_{5/2}$ single neutron
energies for \ca and \bec and in order to get the measured
two-neutron separation energy in  \li and \bea for 1p$_{1/2}$ and 2s states
which are not experimentally well known . Their numerical values can be found
in refs.[1] and [10]. For higher neutron states the
particle-phonon couplings are weak and we take $\alpha_n$=0. We use a
discretisation of the continuum states with a radial box of 20 fm and
orthonormalise the  wave
functions by the Schmidt method. In our previous
papers the two-body term ${\bf p}_1.{\bf p}_2$ was neglected but as the
approximation was the same for all nuclei: $^{14}$C, $^{12}$Be, $^{11}$Li and
$^{14}$Be and the effective neutron-neutron interaction fitted to describe
the properties of $^{14}$C and $^{12}$Be, then used in the other
nuclei, the effect of such an approximation was minimised. 
We have checked that adding this
term changes slightly the strength of the effective interaction  (by few \%)
but gives the same agreements and predictions than previously.  

We used in refs.[1,10] a zero range  density dependent neutron-neutron 
interaction given by:
\be
V_{nn}(1,2)=-V_0\;\left (1-x\;[{\rho_c( { \frac{{\bf r_1+r_2}}{2}})}/
{\rho_0}]^p\right )\delta({\bf r_1-r_2}) 
\ee
where $\rho_c$ is the core density and $\rho_0$=0.16fm$^{-3}$. 
The above fit gives now $V_0$=880 MeV.fm$^{3}$,
 x=0.93 and p=1.2  close to what is used in the
literature \cite{be,tf}. With the same force we
have calculated $^{6}$He as an alpha-particle plus two neutrons. Taking
$V_{nc}$ as the neutron-alpha particle potential fitted 
by Satchler et al. \cite{sao} to low energy l=1 phase shifts we get a
two-neutron separation energy of about 4 MeV in $^{6}$He while it is
 0.97 MeV experimentally \cite{wa}. Similar strong binding was 
 found by Esbensen et al. \cite{hb} 
who
modify the n-n interaction in order to get the experimental value. However
this is not satisfactory for us because we always require that our effective
interaction leads to  an overall agreement for all systems since the density
dependent term should take account of the dependence of the effective 
interaction upon the nucleus.
We have looked for different possible sources of
such unexpected disagreement: choice of neutron-alpha particle interaction,
discretisation of the continuum which will be discussed in a forthcoming paper
and effect of the cut-off on neutron energy. 
However none of these effects is responsible for such a bad result.

The force of eq.(7) has three parameters which are not independently
determined from a fit of energy spectra.
In a recent paper Garrido et al. \cite{gs} have fitted the
three
parameters of $V_{nn}$ in order to get the same gap, $\Delta(k_F)$,   in
nuclear matter than obtained with a Gogny finite range 
effective interaction \cite{gb,jf}. They have shown that to get 
agreement over all the 
domain of $k_F$ they have to take p=0.47, x=0.45 with a
cut-off, $\epsilon_c^0$, on the neutron energy of 50-60 MeV. 
The cut-off energy determines
the strength $V_0$ if one assumes that the density independent part of the
interaction should reproduce the properties of a free neutron-neutron
system. It gives the following relation between $V_0$, $\epsilon^0_{c}$ 
 and  the neutron-neutron scattering length $a_{nn}$ \cite{hb} :
\ba
V_0&=&2{\pi}^2\;\frac{\hbar^2}{m}\;\frac{1}{k_c^{(0)}-{\ds \frac
    {\pi}{2a_{nn}}}}\\
{k_c^{(0)}}^2&=& \frac{2m\epsilon_c^0}{\hbar^2}
\ea
 $a_{nn}$ is experimentally -18.5 fm then very large. If we replace it by 
 -$\infty$ and take $\epsilon^0_c$=60 MeV this relation gives
$V_0$ = 480 MeV.fm$^{3}$ as used in ref.[6]. 

For free neutrons, the neutron energy
is only kinetic energy while in finite nuclei the neutrons are in the potential
$V_{nc}$.
Therefore when using eqs.(8-9) the cut-off energy has to be calculated from the
bottom of the single-particle well and $\epsilon^0_c$ replaced by :
\be
\epsilon^0_c=\epsilon_c+V_{NZ}
\ee
where $V_{NZ}$ is the depth of our one-body potential, eq.(6),
and $\epsilon_c$ the cut-off energy for a neutron described by the
hamiltonian of eq.(3).
 In our calculations we take neutron states up to an
energy $\epsilon_c \simeq$
   10 MeV for all nuclei. $V_{NZ}$ depends on proton and
neutron numbers in the core and varies between about 40 MeV for \li
and 50 MeV for \ca, giving an
equivalent cut-off in free two-neutron system or nuclear matter of 50-60 MeV 
 as required by the fit of ref.[6]. Note that taking $\epsilon_c$ the
same for all nuclei implies that $V_0$ will be slightly different 
for $^{11}$Li, \bec and $^{14}$C.  

This new zero range effective interaction has very different parameters
compared to the usual ones. We have made the same calculations as previously
with this  force for our systems $^{14}$C, \bec and $^{6}$He. \li results 
are not reported in the table but with neutron energies close to
experimental values it is not bound. The results
are reported in Table I where they are compared to the measured two-neutron
separation energy, $S_{2n}$ \cite{wa}. 
We see two interesting facts: the calculated
$S_{2n}$  in $^{6}$He is now 1.02 MeV close to the experimental value but it is
much too low in the other nuclei. 
Since this zero range force is constructed
to reproduce the Gogny force in nuclear matter and in a free two-neutron
system we thought that this equivalence could fail in our finite 
nuclei which are quite far from both systems. 

Then we have made the same Tamm-Dancoff calculation with the genuine 
Gogny force, D1 or D1S \cite{gb,jf}. The Gogny forces are the sum of central
density dependent and spin-orbit terms. For two neutrons coupled to 0$^+$
the spin-orbit term gives very small contribution to pairing matrix elements
and can be neglected \cite{gg}. Therefore we have made our calculations,
keeping the central term only.  The results are very close 
for the two forces and we present  them for the force D1S only.
The results with the Gogny force are presented in Table I.

\begin{center}
\begin{tabular}{|c|c|c|c|} \hline
Force &$^{14}$C &$^{12}$Be & $^{6}$He\\ \hline
Gogny &11.8 & 2.23& 0.94 \\
$\delta$ force & 11.7 & 1.95 & 1.02\\
exp. & 13.12 & 3.67 & 0.97\\ 
\hline
\end{tabular}
\vskip 6mm
TABLE I. Two-neutron separation energies in MeV calculated in Tamm-Dancoff 
model with the Gogny DS1 force and a zaro range force of eq.(7) with 
$V_{0}$=480~MeV.fm$^{3}$, p=0.47 and x=0.45
\end{center}
\vskip 4mm

 We see that they
are the same as obtained with its zero range substitute what tells us that:

1- the equivalence between the two forces, with zero or finite range,  shown
for a free system and nuclear matter holds in all considered finite nuclei.
 We give the results for  $V_0$=480 MeV.fm$^3$, the same for all nuclei, 
 but as already mentionned  $V_0$ should be larger in \bec than in \ca
 following relations (6), (8) and (10)  what will
improve the agreement between the two series of results.

2- the range of the effective force is not responsible for the unability of
the nuclear model to describe simultaneously $^{6}$He and the p-shell nuclei.

The agreement for
$^{6}$He on one side, the disagreement for the other nuclei on the 
other side seem
at first sight a contradiction of the model. However it can 
  be understood when we go back to the first assumption of the Tamm-Dancoff
 model that the core is  inert with neutrons filling the lowest shells. 
 For $^{6}$He the core,
 an
alpha-particle, is strongly bound and very well described in an Hartree-Fock
model with neutrons and protons filling the 1s shell \cite{ca}. 
It means that the
assumption of a closed shell core (implicitly assumed when we put our extra-
neutrons on the unoccupied shells with probability one) is valid in this case. 
However for the other nuclei we know that the cores are not good closed shell 
nuclei. It is known for long in the case of \cc from the work of Cohen-Kurath 
\cite{ck}. For \bec shell model calculations also show a large deviation from 
a pure state as it is required by experiments \cite{so,ba,ss,na}. It is also 
obvious in our calculations for \bec \cite{pvm} or in three-body Fadeev model 
calculations of Thompson et al. \cite{tn} where \bec is described as a core of 
\beb plus two neutrons and found as a large mixture of (2s)$^2$, 
(1p$_{1/2}$)$^2$ and (1d$_{5/2}$)$^2$ two-neutron components while it is 
considered as a pure (2s)$^2$ or (1p$_{1/2}$)$^2$ state when one studies 
$^{14}$Be.

A model which takes account of two-body correlations in the ground 
state of the core is
the particle-particle RPA model more generally called  pair vibration model
 \cite{sa,rp,bvm}. We
briefly recall what  this model is for two neutrons outside a core which, in
an independent particle model, would have closed shells (or subshells) for
neutrons.

\section{Particle-particle RPA model}

Particle-particle RPA relies on the expansion of the two-particle
(two-hole) Green's function in terms of ladder diagrams with upward and
backward going diagrams as shown in Fig.1. Note that a summation over
upward going diagrams only leads to the Tamm-Dancoff approximation of the
previous section. From the related approximate integral equation satisfied
by this RPA two-particle Green's function, a system 
of equations is derived which
describes simultaneously the $A_c$+2 and $A_c$-2 systems where $A_c$
characterises the core which in an independent particle model would be
described as a closed shell (or sub-shell) nucleus.

\begin{figure}
\begin{center}
\begin{tabular}{c}
\epsfysize=5cm
\epsfxsize=10cm
\epsfbox{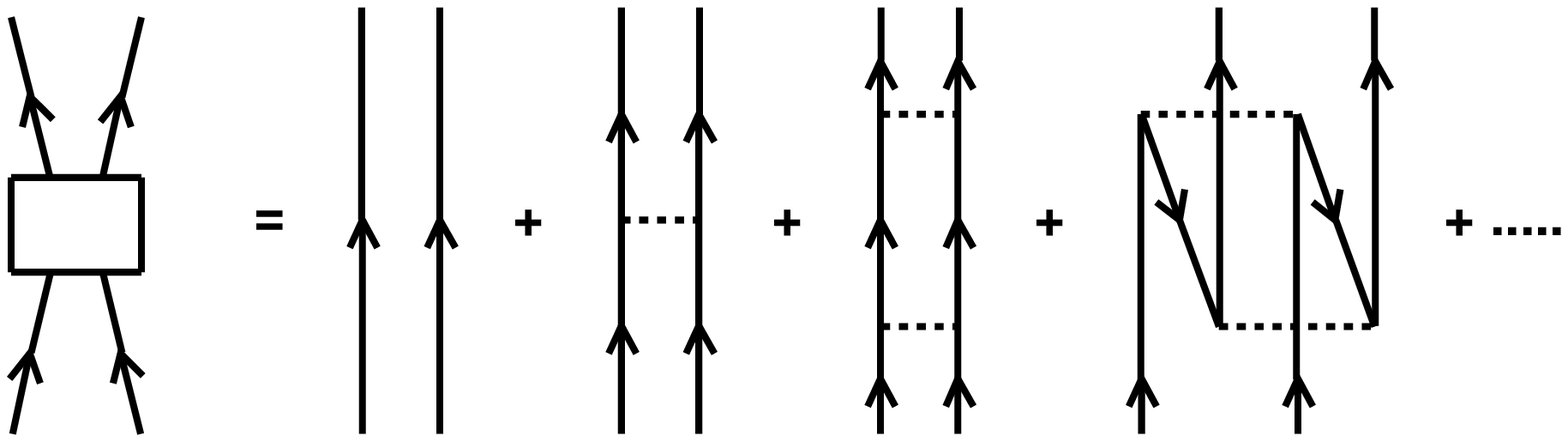}
\end{tabular}
\vskip 1mm
FIG.1: RPA diagrammatic expansion of the two-particle Green's function.
\end{center}
\end{figure}

We here apply this method to our problem of two neutrons outside a core with
$N_c$ neutrons  assuming that the protons are not disturbed when one adds or
subtracts two neutrons.

Let's define a=(a$_1$,a$_2$), b=(b$_1$,b$_2$)$...$ two-neutron configurations 
with the neutrons in states a$_1$,a$_2$$...$ unoccupied in the Hartree Fock 
core ground state, $\alpha$,$\beta$$...$ two-neutron configurations with the 
neutrons in occupied states $\alpha_1$,$\alpha_2$$...$ and two-neutron 
amplitudes or spectroscopic factors as:
\ba
X_{\alpha}(J,M)&=&<N_c+2|A^+_{\alpha}(J,M)|N_c,0>\\
X_a(J,M)&=&<N_c+2|A^+_a(J,M)|N_c,0>
\ea
where the pair operators are given by:
\ba
A^+_a(J,M)&=&\sum_{m_{a_1},m_{a_2}}(j_{a_1},j_{a_2},m_{a_1},m_{a_2}|J,M)
a^+_{a_1}a^+_{a_2}\;\;\;\;\,\,with \,\,a_1\leq a_2\\
A^+_{\alpha}(J,M)&=&\sum_{m_{\alpha_1},m_{\alpha_2}}(j_{\alpha_1},
j_{\alpha_2},m_{\alpha_1},m_{\alpha_2}|J,M)
a^+_{\alpha_1}a^+_{\alpha_2}\;\;\;\;\,\,with\,\,\alpha_1\leq\alpha_2
\ea
$a^+_i$ is the creation operator of a neutron in state $j_i,m_i$, 
$|N_c+2>$ and $|N_c,0>$ are respectively 
the RPA wave functions of the $N_c+2$ nucleus 
in excited or ground state and  of the core in its ground state. 
We see immediately that the amplitudes $X_{\alpha}$ are non zero 
only if the core
ground state has 2p-2h,4p-4h,\ldots
components. Assuming that all $X_\alpha$ are zero, one gets back to the
Tamm-Dancoff approximation.

In the same way we define two-neutron hole amplitudes for the $N_c-2$ nuclei
 from $A_a$ and $A_{\alpha}$, the  annihilation operators for a pair which
 are hermitian 
 conjugates of $A^+_a$ and $A^+_{\alpha}$ defined by eqs.(11-12). They are:
\ba
Y_a(J,M)=<N_c-2|A_a(JM)|N_c,0>\\
Y_{\alpha}(J,M)=<N_c-2|A_\alpha(J,M)|N_c,0>
\ea

For the $N_c$-2 nuclei $Y_a$ are non zero only if the core ground state 
is not a pure HF state but has 2p-2h, 4p-4h\ldots components. Then
these anomalous components $X_{\alpha}$ and $Y_a$ give a measure of the
  two-body correlations in the cores. For a given spin and parity 
 the RPA amplitudes $X$ and $Y$ satisfy the same system of equations:
\ba
(E-\epsilon_a)x_a-\sum   _b<a|V_{nn}|b>x_b-\sum_{\beta}
<a|V_{nn}|\beta>x_{\beta}&=&0\\
(E-\epsilon_{\alpha})x_{\alpha}+\sum_b<\alpha|V_{nn}|b>x_b+\sum_{\beta}
<\alpha|V_{nn}|\beta>x_{\beta}&=&0
\ea
where $x$ are the amplitudes X or Y as explained below.
The two-body matrix elements are antisymmetrised and $\epsilon_a$,
$\epsilon_{\alpha}$
the sum of unperturbed energies of the two neutrons in configurations a,
$\alpha$ respectively.
The one-neutron states are described by the one-body effective potential of
eq.(4) what means that we expand our two-body Green's function of Fig.1 in
terms of  one-body propagators which include particle-phonon
couplings as given in Fig.2. 

\begin{figure}[h]
\begin{center}
\begin{tabular}{c}
\epsfysize=5cm
\epsfxsize=10cm
\epsfbox{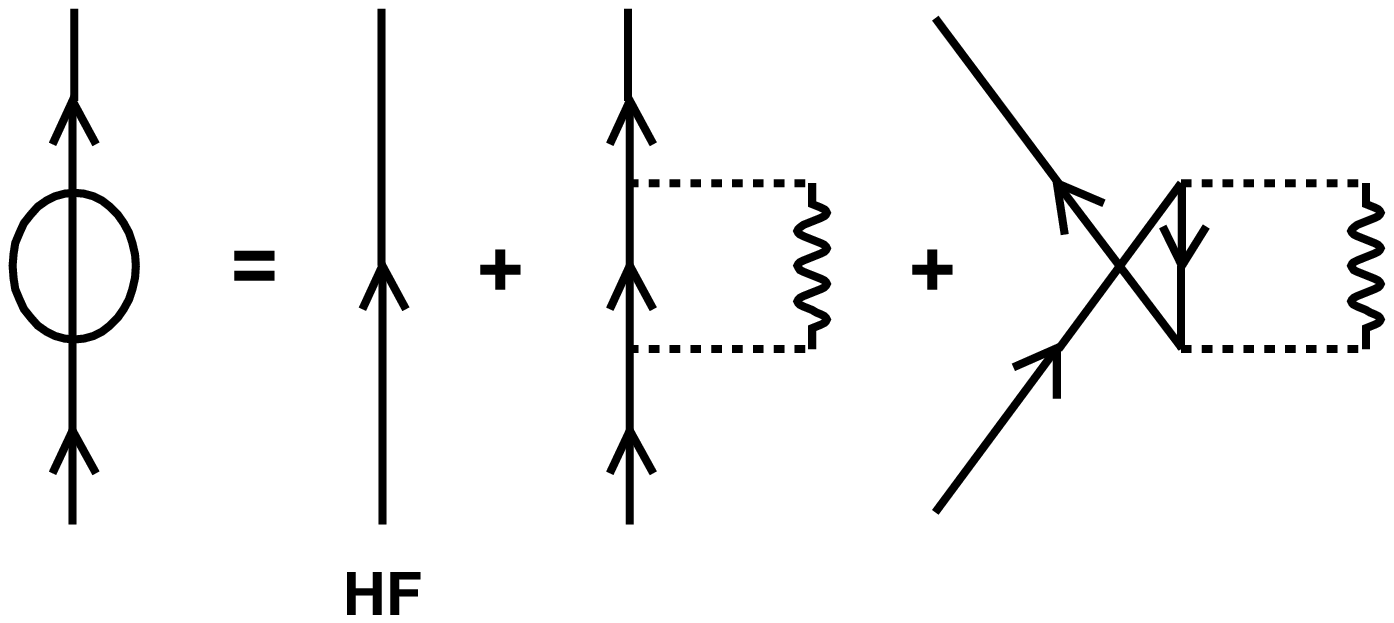}
\end{tabular}
\vskip 1mm

FIGURE 2. One-particle Green's function with inclusion of particle-phonon
          coupling diagrams.
\end{center}
\end{figure}

 If we take N configurations a,b.. and M configurations $\alpha$, $\beta$,.. 
the equations (17,18) have N+M solutions. N solutions correspond to
the $N_c+2$ nucleus with eigenvalues $E_{nJ}$ and amplitudes $X^{(n)}$ 
where n labels the state such that:
\ba
E_{nJ}&=&E_{nJ}(N_c+2)-E_0(N_c)\\
X_a^{(n)}(J)&=&x_a^{(n)}\;\;\;,\;X_\alpha^{(n)}(J)=x_\alpha^{(n)}
\ea
the amplitudes satisfy the following orthonormalisation relation:
\be
\sum_a X_a^{(n)}\;X_a^{(n')}-\sum_\alpha X_{\alpha}^{(n)}\;X_\alpha^{(n')}=
\delta_{nn'}
\ee
The M other solutions (labelled by the index m) are eigenstates 
of the $N_c-2$ nucleus with
eigenvalues $E_{mJ}$ and amplitudes $Y$ given by:
\ba
{\cal E}_{mJ}&=&(E_{mJ}(N_c-2)-E_0(N_c))=-E_{mJ}\\
Y_a^{(m)}(J)&=&x_a^{(m)}\;\;\;,\;Y_\alpha^{(m)}(J)=x_\alpha^{(m)}
\ea
with the orthonormalisation condition:
\be
\sum_a Y_a^{(m)}\;Y_a^{(m')}-\sum_\alpha Y_{\alpha}^{(m)}\;Y_\alpha^{(m')}
=-\delta_{mm'}
\ee
We see from eqs.(19) and (22) that the lowest energies 
$E_{00}$ and ${\cal E}_{00}$
give the two-neutron separation energy in the $N_c+2$ and $N_c$ nuclei
respectively with the relations:
\ba
S_{2n}(N_c+2)&=&-E_{00}\\
S_{2n}(N_c)&=&{\cal E}_{00}
\ea

Then, contrary to particle-hole RPA, all solutions of eqs.(17,18) 
correspond to
physical states. The separation between N and M states follows from the
relative importance of $x_a$ and $x_\alpha$ amplitudes: for the $N_c+2$
nucleus the amplitudes $x_a$ are larger than $x_\alpha$ and inversely for
$N_c-2$ nucleus. 

This model is now applied to  the p-shell nuclei. As noticed already
$^{6}$He described as an $\alpha $ particle plus two neutrons is very likely
well reproduced by a Tamm-Dancoff model since the core is a good Hartree-Fock
system. On the other hand, the $N_c$-2 nucleus (di-proton) is not a bound
system and the RPA model is not reliable in this case so that we discuss the
p-shell nuclei only. 

\section{Results for N=8 nuclei: $^{14}$C, \bec and \li}

We make the calculation for the two effective interactions, the Gogny finite
range interaction and the zero range interaction of eq.(7)
with the parameters p and x
fitted by Garrido et al. and a strength $V_0$ fitted to give the same RPA
energies as the Gogny force. Afterwards we shall  compare
with the strength given by eqs.(8-10).
 The two-neutron subspace (a,b\ldots) is the same as in  section II with a
 normal ordering of 1p$_{1/2}$-2s shells in C-isotopes  but an inversion
 of these two shells   in Be and Li.
 For the   occupied neutron states ($\alpha$, $\beta$\ldots )
 we take the two shells 1s$_{1/2}$ and 1p$_{3/2}$
. The 1p$_{3/2}$ neutron energy   is taken as  the known one-neutron
separation energy for the corresponding core (\cc or \beb or \lic)\cite{wa} .
For the 1s shell we have no
experimental information and we take it as a parameter by changing the depth
of our Saxon-Woods potential. This very deep state
has very little effect on the $N_c$+2 states but a non negligible effect on
the energy of the $N_c$-2
ground states. Then we take the 1s energy  in order to get close agreement, 
if possible, with the experimental two-neutron separation energy
in the core nucleus. This procedure gives $\epsilon$(1s)=-32, -32 and -28
 for $^{12}$C, $^{10}$Be and $^{9}$Li respectively.


\begin{center}
\begin{tabular}{|c|cc|cc|cc|}
\hline
  & $^{14}$C & $^{12}$C & $^{12}$Be &$^{10}$Be & $^{11}$Li & $^{9}$Li \\ \hline
Gogny &12.9 & 32. & 3.69& 8.2 & 0.37 & 3.8  \\
$\delta$ force & 12.9 & 32.5 & 3.6 & 8.52 & 0.34 & 3.76\\
exp. & 13.12 & 31.8 & 3.67 & 8.48 & 0.34 $\pm $0.05 &6.1\\
\hline
\end{tabular}
\vskip 6mm
TABLE II. Same as Table I for RPA model and $V_0$=500, 510 and 560~MeV.fm$^{3}$ 
for $^{14}$C-$^{12}$C, $^{12}$Be-$^{10}$Be and $^{11}$Li-$^{9}$Li respectively
\end{center}
\vskip 4mm

In Table II are presented $S_{2n}$, the two-neutron separation energies 
in $N_c$+2 ($^{14}$C, $^{12}$Be and $^{11}$Li) and $N_c$ ($^{12}$C, $^{10}$Be 
and $^{9}$Li) nuclei given by the lowest eigenvalues $E_{00}$ and 
${\cal E}_{00}$ respectively, following relations (25) and (26). The results 
are given for the Gogny force
and the equivalent zero range force. In Table III we
give the amplitudes $X_a$ and $X_{\alpha}$ for $N_c+2$ nuclei and $Y_a$ and
$Y_{\alpha}$ for $N_c-2$ nuclei corresponding to the ground states and
obtained with the Gogny interaction. The zero range interaction gives nearly
identical eigenvectors.

For $^{14}$C-$^{12}$C and $^{12}$Be-$^{10}$Be, $S_{2n}$ is close to the 
experimental values while we have no free parameters, apart from the 
1s-energy which turns out from the fitting procedure to be very close to a 
typical Hartree-Fock energy. We see that the amplitudes $X_{\alpha}$ 
for \ca and
\bec  are for $\alpha=(1p_{3/2})^2$ rather large  and larger in \bec
than in $^{14}$C. This large value means that in the ground state of the cores, 
$^{12}$C and $^{10}$Be, there are large components of 2nh-2np states with at
least  two
holes in the $1p_{3/2}$ shell. This is in complete agreement with the shell
model results  for \cc \cite{ck} and for \bec \cite{so,ba,ss,na}. 
Furthermore we see that, if in \ca
the amplitude for the two neutrons in a $1p_{1/2}$ state is very large, in   
\bec we have comparable amplitudes for $(2s)^2$, $(1p_{1/2})^2$ and
$(1d_{5/2})^2$ configurations. This means that, together with the fact that
$X_{\alpha}=0.46$, \bec cannot reasonably be considered as a closed shell
nucleus as done in many papers on $^{14}$Be.

Our RPA equations give simultaneously the amplitudes $Y_{a}$ and
$Y_{\alpha}$ of eqs.(15-16) for \cb and $^{8}$Be. Here the $Y_a$ are the 
anomalous amplitudes reflecting again correlations in the cores. We see 
consistency with the results for $^{14}$C and $^{12}$Be. The amplitudes $Y_a$ 
are larger in $^{8}$Be than in \cb with a distribution over several two-neutron 
states revealing a very complicated structure of $^{8}$Be. 

The $^{11}$Li-$^{7}$Li systems present more ambiguity than the
previous ones. Now we know from break-up experiments that the lowest
neutron resonance in $^{10}$Li is an s-state  at 0.1-0.2 
MeV \cite{pf} as required by several calculations \cite{pvm,tz,bh} 
and that the next state is a p$_{1/2}$ resonance  at
0.54$\pm$ 0.06 MeV \cite{li}.

\begin{center}
\begin{tabular}{|c|cccc|cc|}\hline
 &\multicolumn{4}{c|}{$X_a(Y_a)$} &\multicolumn{2}{c|}{$X_\alpha
   (Y_{\alpha})$}\\ \hline
 & (2s)$^2$ & (1p$_{1/2})^2$ & (1p$_{1/2}$,2p$_{1/2}$) & (1d$_{5/2})^2$ &
 (1p$_{3/2})^2$ & (1s)$^2$ \\ \hline 
$^{14}$C &-0.12&0.96&-0.05&-0.28&0.19&-0.09\\
$^{10}$C &-0.05&0.15&-0.02&-0.14&0.98&-0.32\\ \hline
$^{12}$Be & -0.48&0.76&-0.29&-0.44&0.60&-0.09\\
$^{8}$Be  & -0.16&0.34&-0.17&-0.30&1.18&-0.12\\ \hline
$^{11}$Li & 0.66&-0.56&0.48&0.04&-0.45&0.07\\
$^{7}$Li & -0.12&0.22&-0.27&0.02&1.12&-0.11\\
\hline
\end{tabular}
\vskip 6mm
TABLE III. The most important RPA amplitudes ($X_a,X_\alpha $ for $N_c+2$ 
  nuclei, $Y_a ,Y_\alpha$  for $N_c-2$ nuclei)
\end{center}
\vskip 4mm

The results, presented in Tables II and III, have been obtained with
$\epsilon(1s)=-28$ MeV, $\epsilon(2s)=0.19$MeV and $\epsilon(p_{1/2})=0.6$
MeV close to measurements for the two last ones and reasonable for
the first one.      We first see that $S_{2n}$ in \li is
in good agreement with the experimental value, 0.34$\pm$0.05 MeV \cite{lia}. 
The anomalous 
amplitude $X_{\alpha}$ for $\alpha$=(1p$_{3/2}$)$^2$ is as large as it was
in \bec revealing strong deviation from closed shell in $^{9}$Li. We also see
that the amplitudes for (1p$_{1/2}$)$^2$ 
and (2s)$^2$ configurations are
similar, as required by the analysis of break-up experiments \cite{br}.
However 
the $S_{2n}$ value for $^{9}$Li is very far from the measured value, 3.85 instead
of 6.1 MeV. This disagreement, found only in Li, cannot be improved by
a reasonable change of  $\epsilon$(1s) which is the only parameter of the
calculation. It could come from the single last proton which 
is on the same shell as the active neutrons in $^{7}$Li and \lic but we did
not find any simple way to evaluate such an effect.    

 In the case of Li isotopes, there is an ambiguity coming from higher neutron
states. Experimentally one has not seen the d$_{5/2}$ resonance, therefore we
have calculated it with the bare Woods-Saxon potential even though
we know that in
$^{11}$Be one has to add a surface term which lowers the d$_{5/2}$ energy.
 However the effect of such a term should not affect the results since, as
 seen in Table III, the amplitude for the (1d$_{5/2})^2$ configuration is
 very small. Also, because  $^{10}$Li
is unbound,  all neutron states are in the continuum with the result
that continuum non-resonant states play a more important role in the
determination of the \li energy. This effect is
particularly important for the discretised 2p$_{1/2}$ state which is
taken as an
eigenstate of the bare Woods-Saxon potential. Because the modification of
the potential by particle-vibration couplings is very large for the
1p$_{1/2}$ resonance, this 2p$_{1/2}$ state comes at an energy  close to
the 1p$_{1/2}$, giving a large overlap between the two states. We know that 
particle-vibration
 couplings are more important for states close to the Fermi
surface and can be neglected for higher states in usual nuclei. However in
$^{10}$Li these couplings are very strong and very likely still efficient
for the 2p$_{1/2}$ state. Then as a test we have made the calculation with
the same surface contribution to the
one body potential for
1p$_{1/2}$ and 2p$_{1/2}$ states, taking the bare potential for the other
p-states which anyway have much higher energies and therefore have a weaker
effect on the ground state energy and wave function. The amplitude $X_a$ for
a=(1p$_{1/2}$,2p$_{1/2}$) is weaker, $S_{2n}$ slightly smaller. To recover 
  the same value  one has to take
  $\epsilon$(1p$_{1/2}$)=0.53 MeV, then at  the experimental energy,  but
  these differences are 
   not significant. 

Going back to Table II we see that 
the strengths $V_0$ of the zero range interaction fitted to recover the same
energy than the Gogny interaction are 500, 510 and 560 MeV.fm$^{3}$ for 
$^{14}$C, $^{12}$Be and $^{11}$Li respectively. They follow closely 
the dependence on $V_{NZ}$ given by 
 eqs.(8-10) with a value for \ca very close to  480 MeV.fm$^{3}$ used
 in ref.[6].   We again find the same equivalence 
 between the two forces,with zero or finite range, as 
derived for free neutrons and nuclear matter. It seems from our study that
the need of a much stronger zero range force to get experimental binding
energies in  simple pairing models, as reminded in Section II, is in
fact a way to take implicitly  account of two-body correlations in the core 
which   are neglected in the model.

\section{Results for \bea (N=10).}

The problem of \bea is not yet completely clarified, both experimentally and
theoretically. Experimentally one knows that $^{13}$Be has a
d$_{5/2}$ resonance at 2.01 MeV
above the n+$^{12}$Be threshold \cite{ost}, a lower resonance at about
0.8 MeV seen in the reaction $^{14}$C($^{11}$B,$^{12}$N)$^{13}$Be \cite{pos}
which has no spin or parity assignment and, from a recent experiment \cite{th}
 using fragmentation of
$^{18}$O and detecting neutrons in coincidence with $^{12}$Be, a 1/2$^+$
resonance below 0.2 MeV which should be  the ground state of this
unbound nucleus. Theoretical models describing \bea as two neutrons
outside a core of $^{12}$Be, where the neutrons fill the 1s, 1p$_{3/2}$ and
1p$_{1/2}$ shells, have either to lower the d$_{5/2}$ resonance or to assume
a bound 2s state therefore to bind $^{13}$Be to get a correct binding energy
in \bea \cite{tn}. In a first paper \cite{sl}
based on a two-neutron Tamm-Dancoff model and with a zero range effective
interaction fitted on \ca  and \bec we found that an inversion of the
1p$_{1/2}$-2s shells leads to  the correct binding
energy in \bea without modifying the known d$_{5/2}$ energy nor assuming a
bound
$^{13}$Be. This calculation, as others, assumes a closed shell nucleus of
\bec where the neutrons fill the 1s-1p$_{3/2}$-2s shells, 
while we have seen above that the ground state of \bec has amplitudes
$X_a$ 
over several two-neutron configurations and, because of large $X_\alpha$, has
very likely components on more complicated
configurations. The
particle-particle RPA  applied to \bea takes this into account 
and will provide a description 
of the core consistent with the RPA amplitudes of Table III.

We have made the RPA calculation assuming a normal ordering of shells:
1s, 1p$_{3/2}$, 1p$_{1/2}$, 2s,\ldots with a Hartree Fock state where the p-shell
is filled for neutrons. With $\epsilon$(1p$_{1/2}$)=-3.15 MeV, given by the
neutron separation energy in $^{12}$Be \cite{wa}, a 1d$_{5/2}$ state at 2 MeV,
the
experimental value, and a 2s state at 7 keV,  very close to the threshold, 
we get S$_{2n}$=0.7 MeV much too low compared to the experimental value of
1.34$\pm$0.11 MeV \cite{wa}. For \bec we get S$_{2n}$=3.26 MeV, also 
too small.  Moreover the amplitudes $X_{\alpha}$ are small indicating
weak correlations in \bec in disagreement with the results of the previous
section. Therefore we find again that we are not able  to
describe satisfactorly \bea when we assume a normal ordering of 
shells in $^{12}$Be.

We now assume an inversion of the two shells 2s and 1p$_{1/2}$ as in $^{11}$Be.
 The configurations $\alpha$ are built on 1s,1p$_{3/2}$ and 2s states while
 the configurations a include neutrons in 1p$_{1/2}$ state. 
The results are summarised in Table IV where we give S$_{2n}$ for
\bea and the RPA amplitudes for $^{14}$Be and $^{10}$Be. They are
obtained for a Gogny effective interaction with the d$_{5/2}$ resonance at
the experimental energy of 2 MeV, a 1p$_{1/2}$ resonance at 0.68 MeV, an
 occupied 2s shell with an energy of -3.15 MeV given by 
 the experimental neutron
 separation energy in \bec and a 1p$_{3/2}$ state at -5.6 MeV. 
 For the $\alpha$ configurations we have taken the
 (1p$_{3/2}$)$^2$, (1s,2s) and (2s)$^2$ states. 
  We have checked that adding the (1s)$^2$
 states does not change our results. 

 We get 3.65 MeV for the two-neutron separation energy in $^{12}$Be. By 
comparing with the result of Table II we see  that the values 
of S$_{2n}$ in \bec found in both calculations
are very close showing the coherence of the model. Indeed we get 3.65 MeV when 
\bec is considered as the core for the $^{14}$Be-$^{10}$Be systems while it is
3.69 MeV when it is described as two neutrons outside a core of $^{10}$Be.

The p$_{3/2}$ energy cannot be deduced directly from experiments.
However $^{11}$Be has a 3/2$^-$ state at 3.9 Mev excitation energy \cite{aj},
then at 7 MeV when referred to the \bec core, with a very small width and has
certainly a large component of one neutron-hole in $^{12}$Be 
 mixed with  
2h-1p components which have higher energies. Therefore  a hole energy of
5.6 MeV is not unrealistic. 

According to our discussion for $^{11}$Li we have done the calculation 
using  the same surface potential for 1p$_{1/2}$ and 2p$_{1/2}$ states. We
have then to lower the 1p$_{1/2}$ energy to 0.56 MeV if we leave all other states
to be the same , what does not modify qualitatively our conclusions.


\begin{center}
\begin{tabular}{|c|c|ccc|cc|}\hline
 & S$_{2n}$(MeV) & (1p$_{1/2})^2$ & (1p$_{1/2}$,2p$_{1/2}$)& (1d$_{5/2})^2$ &
 (1p$_{3/2})^2$ & (2s)$^2$ \\ \hline
$^{14}$Be & 1.30 & -0.46 & 0.62 & 0.74 & -0.64 & 0.46 \\
$^{10}$Be & - & -0.20 & 0.31 &0.46 & -0.73 & 1.\\
\hline
\end{tabular}
\vskip 6mm
TABLE IV. RPA energy in MeV for \bea and main amplitudes for \bea and \beb
  obtained with the D1S Gogny force.
\end{center}

\vskip 4mm


The amplitudes given in Table IV show again strong correlations in $^{12}$Be.
 Indeed the
anomalous amplitudes in \bea are 0.64 and 0.46  for (p$_{3/2})^2$ and (2s)$^2$
respectively what means that in \bec components on configurations with two
holes on these  shells are important. Note that the components with two holes
in the 2s-shell are qualitatively consistent with the
amplitudes $X_a$ of Table III for a=(1p$_{1/2}$)$^2$ and (1d$_{5/2})^2$.
However RPA gives only overlap of wave functions and a strict and direct
comparison between amplitudes derived in the two calculations is not
possible. To make a direct comparison one has to calculate wave functions
what can be made using  a quasi-boson approximation as was done in the past for
particle-hole RPA correlations.

Our fitted value $\epsilon(1p_{1/2})$=0.68 MeV is consistent with an unbound
1/2$^-$ state in $^{13}$Be which would be at 0.68 MeV above the \bec+n
threshold, close to the experimental resonance at 0.8 MeV. However the
recent experiment using fragmentation of $^{18}$O \cite{th} shows
 a low energy ($\leq$0.2 MeV) s-wave strength in $^{13}$Be what,  in
 an independent
particle model,  would mean that the lowest unoccupied shell in \bec is an s
shell and would reject the possibility of inversion. In our model however
$^{13}$Be is described as a neutron added to the correlated core of
$^{12}$Be which is a mixture of many different states, in particular
it has large components on configurations with
 a closed 1p$_{3/2}$ shell plus two neutrons on the 2s or on the 1p$_{1/2}$
shell. Therefore these two components will give in $^{13}$Be two different
states, a 1/2$^-$ state built on the first one with the last
neutron on the p$_{1/2}$ shell (the 2s-shell is filled)
and a 1/2$^+$ state built
on the second one with the last neutron on the empty 2s
shell. In a weak  coupling model, because of the known inversion in
n+$^{10}$Be system, one can show that the 1/2$^+$ state is
lower than the  1/2$^-$ state by about 0.32 MeV and at about 0.3 MeV
above the $^{12}$Be+n threshold what is in qualitative agreement
with the  experimental state seen recently. Consequently 
a 1/2$^+$ ground state in $^{13}$Be is not in contradiction 
with an inversion of the 2s-1p$_{1/2}$ shells.

There are several other arguments in favor of this inversion in 
$^{12}$Be-$^{13}$Be. The first one relies on the recent measurement of 
the B(E2) for the 2$^+$ state at 2.1 MeV in \bec which is found to be 
the same as in $^{10}$Be for the 2$^+$ state at 3.4 MeV \cite{bei}. Because 
the inversion in $^{11}$Be is related to the large value of the B(E2) in 
$^{10}$Be \cite{nvm,nun} there is no reason why the effect 
should be smaller in $^{13}$Be. Moreover the phonon has a smaller energy  in
\bec than in $^{10}$Be what is expected to give an 
enhancement of the coupling \cite{nvm}.
Note that this large B(E2) was predicted in ref.[10].
Further arguments are found by looking at the $^{11}$Be spectrum. Indeed if
$^{13}$Be can be described as a neutron added to a core of $^{12}$Be, $^{11}$Be
can be described as a hole in the same core of $^{12}$Be. Therefore the 1/2$^+$
ground state of $^{11}$Be is expected to  correspond mainly to a hole in 
the last occupied shell in the Hartree-Fock ground state of \bec what implies 
that this shell should be an s shell, not a p$_{1/2}$.
The first excited states   can be obtained as two holes coupled
to 0$^+$ plus a neutron on the p$_{1/2}$ or the d$_{5/2}$ shell 
therefore will be a 1/2$^-$ and a higher 5/2$^+$ states respectively, 
in agreement with
the experimental spectrum. If we assume that the residual interaction
between the two holes and  the particle are similar if the particle is on
a p$_{1/2}$ or d$_{5/2}$ shell we find a difference between the excitation
energies of the two states of 1.32 MeV while experimentally it is 1.45 MeV.
One sees that the  inversion is able to give a coherent description of 
$^{11}$Be, $^{13}$Be and \bea.

Because any model relies on approximations one should be aware of the
difficulty to draw a precise scheme. For this reason it is desirable to
compare results of different models. Our results without inversion of shells
agree with those of Thompson and Zukhov \cite{tn}.   We now
make a comparison with the work of
Descouvemont \cite{deb,de}. He has calculated in the generator
coordinate model (GCM) $^{13}$Be
and \bea as \bec+n and \bec+n+n systems respectively. The  core of \bec is
described by a filled 1p-shell for the neutrons while the two protons of the
1p-shell can couple to different states, ground and excited 0$^+$, 1$^+$ and
2$^+$  states. Qualitatively it is equivalent   to our Tamm-Dancoff approach
where contribution  of core excited states are put in our one-body neutron
potential and where the neutrons of the core are assumed to fill the 1p-shell.
 The GCM calculations lead to a slightly bound 1/2$^+$ ground state
 for $^{13}$Be and a
 \bea bound by 1.1 MeV. In our Tamm-Dancoff approach with the zero range
 force fitted on $^{14}$C and $^{12}$Be, a 2s neutron state at -90 keV and a
filled neutron 1p-shell (therefore without inversion) we get for \bea a
binding energy of -0.93 MeV what is close to the GCM result. The three
models, three-body Faddeev, GCM or simple pairing, lead to similar results
but are not able to give rise to good agreement for
$^{13}$Be and \bea when in \bec the neutrons are assumed to fill the
1p$_{3/2}$-1p$_{1/2}$ shells. 

A different work \cite{ma} using a density dependent relativistic mean field 
 model calculates 
$^{12}$Be-$^{14}$Be assuming closed 1p and 1p-2s neutron shells 
respectively. It gives too large one-
neutron and two-neutron separation energies in both systems. This result, 
together with the assumption of closed shells in both $^{12}$Be and $^{14}$Be, 
is not in agreement with other calculations.  

\section{Conclusions}

We have first shown that a two-neutron Tamm-Dancoff model with a zero range
density dependent neutron-neutron interaction fitted on \ca and \bec gives
simultaneously good results for \li and \bea but fails to describe $^{6}$He. 
The zero range force necessary to get agreement in C-Be-Li nuclei has very 
different 
parameters compared to the parameters fitted by Garrido et al. to reproduce
the gap calculated in nuclear matter with the finite range Gogny effective
interaction. The same Tamm-Dancoff calculation with these two forces shows
that they are still equivalent in our finite nuclei and gives a
good binding energy in $^{6}$He but too weak binding in N=8 nuclei.
This result is well understood in terms of two-body correlations in the
cores. Indeed we know that the alpha-particle is well described in
Hartree-Fock model. Then two-body correlations in $^{4}$He are unefficient
while we know for long from shell model calculations that the cores of \cc
and \bec and very likely $^{9}$Li are not pure closed shell nuclei as assumed in
a pairing model. It is also obvious for \bec in our calculation: when it is
described as a core of \beb plus two neutrons its wave function is a mixture
of (2s)$^2$, (1p$_{1/2}$)$^2$ and (1d$_{5/2}$)$^2$ two-neutron states while
in the study of \bea it is considered as a pure (2s)$^2$  or
(1p$_{1/2}$)$^2$ state what yields inconsistency of the model.

The particle-particle RPA model is well adapted to take into account such 
correlations and indeed the model applied to 
$^{14}$C-$^{11}$Li-$^{12}$Be-$^{14}$Be gives
now with those realistic forces very good agreement with experimental
binding energies. It gives also the
two-neutron separation energy in the cores. For $^{12}$C-$^{10}$Be and \bec
(the latter being considered
as a core in the calculation of $^{14}$Be) the agreement with measurements is
also
very good. However it is too small in $^{7}$Li  very likely due to the single
proton in the p$_{3/2}$ shell. The model gives also two-neutron 
and two-neutron hole amplitudes in the wave functions (spectroscopic factors)  
which are related to the amount of two-body correlations introduced in the
cores. This effect is always large but larger in $^{10}$Be-$^{12}$Be than in
$^{12}$C, in  qualitative  agreement with shell model calculations.

 To get a good two-neutron separation energy in \bea and to get a consistent
 description of \bec when it is considered as the core of \bea or described
 as \beb + two neutrons, one has to assume an inversion of 2s-1p$_{1/2}$
shells in $^{12}$Be-$^{13}$Be as it is in $^{10}$Li and $^{11}$Be.
 This inversion is also suggested by
 a  recent measurement \cite{bei} of the transition 2$^+$(2.1 MeV)
$\rightarrow$ 0$^+$(gs) in $^{12}$Be. The B(E2) is found to be the same as for the
 transition 2$^+$(3.3 MeV) $\rightarrow$ 0$^+$(gs) in \beb,
suggesting the same effect of particle-phonon couplings in the two systems,
$^{11}$Be-$^{13}$Be, 
therefore the same shells inversion.
 Furthermore strong particle-particle RPA correlations are known to modify the
 one-neutron mass operator \cite{vm,bd} therefore to give further
 corrections to the
one-neutron single energies. They may 
enhance the inversion process studied in refs.[11,34] for $^{11}$Be,
even though when
adding the two contributions due to couplings with phonons and pair
vibrations one has to substract the second order term in order to eliminate
double counting, so that only very mixed RPA states will contribute. From the
calculated amplitudes one may expect this contribution to be non
negligible in  $^{11}$Be and $^{13}$Be and even larger in $^{13}$Be than in
$^{11}$Be.

One of us (NVM) is very grateful to P. Schuck for very fruitful discussions
which have initiated an important part of this work. 
We also are indebted to G.F. Bertsch for
many comments and suggestions during the preparation of the manuscript. 
The second version of the manuscript has been carefully  read by F.
Bortignon, D. Kadi-Hanifi and P. Schuck. We thank them for accepting this
tedious job and for their helpful remarks.


\begin{thebibliography}{50} 
\bibitem{pvm} N. Vinh Mau and J.C. Pacheco, \npb {A607}, 163 (1996).  
\bibitem{wa} G. Audi and A.H. Wapstra, \npb {A565}, 66 (1993).  
\bibitem{hb}  H. Esbensen, G.F. Bertsch and K. Hencken, \pr {56}, 3054 (1997)
\bibitem{ca} X. Campi, J. Martorell and D W.L. Sprung, \plb {41}, 443 
  (1972).  
\bibitem{ck} S. Cohen and D. Kurath, \npb {73}, 1 (1965).  
\bibitem{gs} E. Garrido, P. Sarriguren, E. Moya de Guerra and P. Shuck, \pr
  {60}, 064312 (1999).   
\bibitem{gb} J. Decharge and D. Gogny, \pr {21}, 1568 (1980).  
\bibitem{jf} J.F. Berger, M. Girod and D. Gogny, Comp. Phys. Comm. {\bf 63},
 365 (1991).
\bibitem{be} G.F. Bertsch and H. Esbensen, \apb {209}, 327 (1991).
\bibitem{sl} M. Labiche, F.M. Marques, O. Sorlin and N. Vinh Mau, \pr {60},
 027303 (1999).   
\bibitem{nvm} N. Vinh Mau, \npb {A592}, 33 (1995)
\bibitem{tf} J. Terasaki, H. Flocard, P.H. Heenen and P. Bonche, \npb {A621}, 
 706 (1997).  
\bibitem{sao} R. Satchler, L.W. Owen, A.J. Elwyn, G.L. Morgan and R.L. Walter,
 \npb {A112}, 1 (1968). 
\bibitem{gg} M. Girod and B. Grammaticos, \pr {27}, 2317 (1983). 
\bibitem{so} T. Suzuki and T. Otsuka, \pr {56}, 847 (1997)  
\bibitem{ba} B.A. Brown, International School of Heavy Ion Physics: Exotic
  Nuclei, edited by R.A. Broglia and P.G. Hansen (World Scientific,
  Singapore, 1998) p.1 
\bibitem{ss}H. Sagawa, T. Suzuki, H. Iwasaki and M. Ishihara, \pr
    {63},  034310(2001).
\bibitem{na} A. Navin et al., \prla {85}, 266 (2000). 
\bibitem{tn} I.J. Thompson and M.V. Zukhov, \pr {53}, 708 (1996).  
\bibitem{sa} N. Fukuda, F. Iwamoto and K. Sawada, \praa {135}, 
932 (1964). 
\bibitem{rp} G. Ripka and R. Padjen, \npb {A132}, 489 (1969).
\bibitem{bvm} A. Bouyssy and N. Vinh Mau, \plb {35}, 269 (1971); \npb
  {A224}, 331 (1974).  
\bibitem{pf} S. Pita, These, Universite Paris 6 (2000) 
\bibitem{tz} I.J. Thompson and M.V. Zukhov, \pr {49}, 1904 (1994).  
\bibitem{bh} G.F. Bertsch, K. Hencken and H. Esbensen, \pr {57}, 1366 (1998) 
\bibitem{li} R.M. Young et al., \pr {49}, 279 (1994).
\bibitem{lia} T. Kobayashi, \npb {A538}, 343c (1992).    
\bibitem{br} M. Zinzer et al., \prla {75}, 1719  (1995). 
\bibitem{ost} A.N. Ostrowski et al., \zpb {A343}, 489 (1992).  
\bibitem{pos} A.V. Belozyorov et al., \npb {A636}, 419 (1998).  
\bibitem{th} M. Thoennessen, S. Yokoyama and P.G. Hansen, \pr {63},
 014308 (2000). 
\bibitem{aj} F. Ajzenberg-Selove, \npb {A506}, 1 (1990).   
\bibitem{bei} H. Iwasaki et al., \plb  {481}, 77 (2000).
\bibitem{nun}F.M. Nunes et al., \npb {A609}, 43 (1996).    
\bibitem{deb} P. Descouvemont, \plb  {331}, 271 (1994).
\bibitem{de} P. Descouvemont, \pr {52}, 704 (1995). 
\bibitem{ma} Zhongzhou Ren, Gongou Xu, Baoqiu Chen, Zhongyu Ma and W. Mittig 
\plb {351}, 11 (1995).  
 \bibitem{vm} N. Vinh Mau, Theory of nuclear structure: Trieste Lectures
   1969,(IAEA Vienna) (1970) p.931
\bibitem{bd} C. Barbieri and W.H. Dickhoff, \pr {63}, 034313 (2001). 
\end{thebibliography}
\end{document}